\documentclass{raa}            

\usepackage{graphicx,times}             
\usepackage{natbib}
\usepackage{pdflscape}  
\usepackage{float}  
\usepackage{amssymb,amsmath}
\usepackage{makecell}
\bibpunct{(}{)}{;}{a}{}{,}

\usepackage[pagebackref=true]{hyperref}
\usepackage{lineno}
\usepackage{longtable}
\usepackage{array}
\usepackage{caption}
\begin{document}

  \title{Searching for Spider-Like Pulsars from TESS Ellipsoidal Lightcurves with X-ray counterparts
}

   \volnopage{Vol.0 (20xx) No.0, 000--000}      
   \setcounter{page}{1}          

   \author{Xiaoqing Liang
   \inst{1,2}
   \and Partha Sarathi Pal
   \inst{3,4}
   \and P.~H.~T.~Tam
   \inst{1,2}
   \and Rishank Diwan
   \inst{3}
   \and Wen-Jun Huang
   \inst{1,2}
}

\institute{
   School of Physics and Astronomy, Sun Yat-Sen University, Zhuhai 519082, People's Republic of China; {\it tanbxuan@sysu.edu.cn}\\
   \and
   CSST Science Center for the Guangdong-Hongkong-Macau Greater Bay Area, Sun Yat-Sen University, Zhuhai 519082, People's Republic of China\\
   \and
   Laboratory for Space Research, The University of Hong Kong, Room 405B, 4/F, Block A, Cyberport 4, 100 Cyberport Road, Hong Kong SAR\\
   \and
   Institute of Astronomy Space and Earth Science, P177, CIT Road, Scheme-7M, Kolkata, 700054, West Bengal, India\\
\vs\no
   {\small Received 2025 April 7; accepted 2025 December 22}}

\abstract{ We present a search for new spider pulsar candidates through multi-wavelength cross-matching, including $\gamma$-ray, X-ray, and optical data. A search for sinusoidal-like optical modulations in TESS data of 183 eROSITA X-ray sources coincident with unassociated \textit{Fermi}-LAT $\gamma$-ray sources led to the identification of four promising spider pulsar candidates. We found optical variability periods ranging from 5 to 13 hours. All candidates display smooth sinusoidal-like phase light curves, similar to what can be expected from ellipsoidal variation; one shows double-peaked profiles indicative of harmonics. The absence of sharp minima, which are often found in black widow systems due to irradiation, together with their optical magnitudes of about G $\approx$ 14, suggests these sources are more likely redback-type binaries. One of the Fermi-LAT counterparts is included in a machine-learning catalog of unassociated $\gamma$-ray sources, with relatively high pulsar probabilities. We also identify potential Gaia counterparts for several sources and estimate their distances and luminosities where parallax measurements are available. Future observations, including further spectroscopic and multi-wavelength studies, are needed to fully characterize these systems.
}
   \authorrunning{}            
   \titlerunning{}  

   \maketitle

\section{Introduction}      \label{sect:intro}
Pulsars are born from remnants of massive stars that have exploded as supernovae. Pulsars utilize their strong magnetic fields and rapid rotation to emit radiation, which depletes their rotational energy. The loss of energy results in a gradual deceleration of the pulsars' rotation. A pulsar is considered 'dead' when its spin-down power is insufficient to sustain the radiation beam. In compact binary systems, dead pulsars are able to accrete material from the companion star. The neutron star will acquire enough angular momentum to reactivate particle acceleration within the magnetosphere, leading to the re-emergence of pulsed radiation. This process marks the formation of a rotation-powered millisecond pulsar (MSP). Recycling is thought to be the main channel through which MSPs are formed \citep{1982Natur.300..728A}.

Spiders are compact X-ray binaries consisting of MSPs and their low-mass companion stars, typically with orbital periods of a few hours. Early work identified such systems through radio eclipses and companion ablation, most notably the discovery of the first “black widow” pulsar PSR B1957+20 in the Galactic field \citep{1988Natur.333..237F}, followed by theoretical studies on the evaporation of very-low-mass companions \citep{1989ApJ...343..292R}. In terms of their evolutionary context, spider systems are thought to arise from low-mass X-ray binaries through stages including sustained mass transfer, possibly common-envelope evolution, followed by loss of angular momentum \citep{1989ApJ...336..507R}. This leads over time (on the order of ~10$^8$–10$^9$ yr \citep{2003A&A...399..237E}) to very short orbital periods, low companion mass, and companion-star heating or ablation by the pulsar wind \citep{2002ApJ...565.1107P,2012ApJ...753L..33B}.  Based on the mass of their companion stars, spiders are classified into two categories, redbacks (RBs) and black widows (BWs). RBs usually have non-degenerate companion stars with masses ranging from $0.1$–$0.4\,M_\odot$. BWs have degenerate companion stars with very low masses, typically less than $0.1\,M_\odot$ \citep{2019Galax...7...93H,2013IAUS..291..127R}. Due to their small intra-binary distances, the orbital periods of spider pulsars are typically only a few hours. The evolutionary mechanisms of spiders remain uncertain. However, there are spider systems with companion stars whose masses fall between those of black widows and redbacks \citep{2023Natur.620..961P}. Since the companion star mass estimates strongly depend on the assumed orbital inclination, these intermediate values may partly reflect measurement uncertainties \citep{2025MNRAS.536.2169S}.  This suggests the possibility that redbacks may evolve into black widows \citep{2015ApJ...814...74J, 2025A&A...698L...5B}.

Most MSPs are found in binary systems, supporting the spin-up evolutionary scenario. However, approximately 20\% of MSPs observed in the Galactic field are isolated. Since MSPs in spider systems ablate their companion stars, the companion stars will eventually vanish if sufficient time has passed \citep{1988Natur.334..227V}. The descendants of some spiders could be isolated MSPs. 

Since the launch of the \textit{Fermi Large Area Telescope} (\textit{Fermi}-LAT) in 2008, it has provided an important new tool for the discovery of pulsar systems. A number of binary systems, including spiders, have been detected and confirmed through multi-wavelength observations. There are 50 confirmed black widows and 30 confirmed redbacks \citep{2025ApJ...994....8K}, and more candidates have been found thereafter. Optical searches can effectively identify potential spider candidates, but their confirmation requires follow-up at multiple wavelengths \citep[e.g.,][]{2024ApJ...977...65T,2025ApJ...978..106L}.
Irradiation and ellipsoidal variation are two main factors that affect optical light curves. Irradiation from the pulsar wind produces asymmetric orbital light curves, typically characterized by a broad maximum and a sharp minimum near superior conjunction \citep{2001ApJ...548L.183S}, and also induces orbital color changes due to temperature differences between the heated and unheated sides of the companion, which can be diagnosed with multi-band photometry. In contrast, ellipsoidal variation due to tidal distortion leads to a double-peaked, nearly sinusoidal modulation with two maxima and two minima per orbit. 

In this work, we combine TESS photometry, eROSITA X-ray survey data, and Fermi-LAT $\gamma$-ray observations to identify and characterize candidate spider systems. Specifically, we examine TESS optical light curves at the positions of eROSITA sources lying within the \textit{Fermi} 4FGL error ellipses \citep{2023arXiv230712546B}. An overview of the observations and data reduction is given in Section~\ref{sec:obs}. The main package, functions, and methods used in the data analysis are in Section~\ref{sec:analysis}. We report our searching results with sky map, power density spectrum and folded light curve in Section~\ref{sec:result}. Some discussions are presented in Section~\ref{sec:discussion}, and the conclusions are summarized in Section~\ref{sec:conclusion}.

\section{Observations and Data Reduction} \label{sec:obs}
\subsection{Fermi-LAT}
The \textit{Fermi}-LAT 14-year Source Catalog (4FGL-DR4) contains 7,187 $\gamma$-ray sources, with roughly one-third of them being unassociated~\citep{Abdollahi_2020}. To identify potential spider pulsar candidates, we first selected sources from the 4FGL-DR4 based on the following criteria:

\begin{itemize}
\item The source is unassociated, meaning it has no confirmed counterpart at other wavelengths.

\item The detection significance is greater than 5, ensuring a reliable $\gamma$-ray detection.

\item The spectral model is either a log-parabola or a power law with exponential cutoff, which is expected for pulsar-like sources.

\item The variability index is smaller than 27.69, a threshold below which the source is considered non-variable at a 99\% confidence level.
\end{itemize}

These criteria were chosen to preferentially select steady, high-confidence $\gamma$-ray sources without known associations, consistent with the properties of pulsars or compact binary systems such as black widow and redback systems.

\subsection{eROSITA}
The eROSITA (extended ROentgen Survey with an Imaging Telescope Array) mission \citep{2024A&A...682A..34M} is a space-based X-ray observatory that conducts wide-field surveys of the sky in the 0.2–10 keV range. With its high sensitivity and ability to perform deep X-ray observations, eROSITA is a powerful tool for detecting X-ray emission from compact objects such as spider pulsars. The X-ray luminosities of spider systems arise from both intrabinary shocks caused by the interaction of the pulsar wind with the companion star \citep{2020ApJ...904...91V} and from the polar cap emission on the neutron star surface \citep{2011ApJ...730...81B}, make these systems ideal targets for eROSITA's survey.

The eROSITA 1B single-band source catalog includes all detections with a detection likelihood DET LIKE $\ge$ 5 in order to maximize completeness. To ensure the reliability of our sample and minimize contamination from false detections, we restrict our selection to eROSITA sources with DET LIKE $>$ 10, for which the spurious fraction is estimated to be as low as 1\% \citep{2022A&A...665A..78S}.

\subsection{TESS}
The Transiting Exoplanet Survey Satellite (TESS) is a survey satellite with a bandpass of 600--1000 nm, primarily designed to monitor the flux variations of stars to detect exoplanets orbiting them \citep{2015JATIS...1a4003R}. With its exceptional timing precision, capable of capturing variations on timescales ranging from minutes to hours, and its nearly all-sky coverage, TESS is also an excellent instrument for characterizing periodicities and flux variations in stars and binary systems~\citep[e.g.,][]{2020ApJ...895L..36P}. 

\subsection{Cross-matching of Fermi-LAT, eROSITA, and TESS}
We cross-matched the Fermi-LAT and eROSITA catalogs to identify eROSITA X-ray sources located within the 95\% error ellipses of 4FGL sources, and then selected those that fall within the TESS field of view. This process resulted in 183 candidates shown in Table~\ref{mrt}. Among these, we identified four eROSITA candidates whose TESS light curves exhibit periodic variations that coincide with their positions and display sinusoidal-like modulations.

\section{Data Analysis} \label{sec:analysis}
\begin{figure}[h!]
    \centering
    \begin{tabular}{cc} 
        \includegraphics[width=0.48\textwidth]{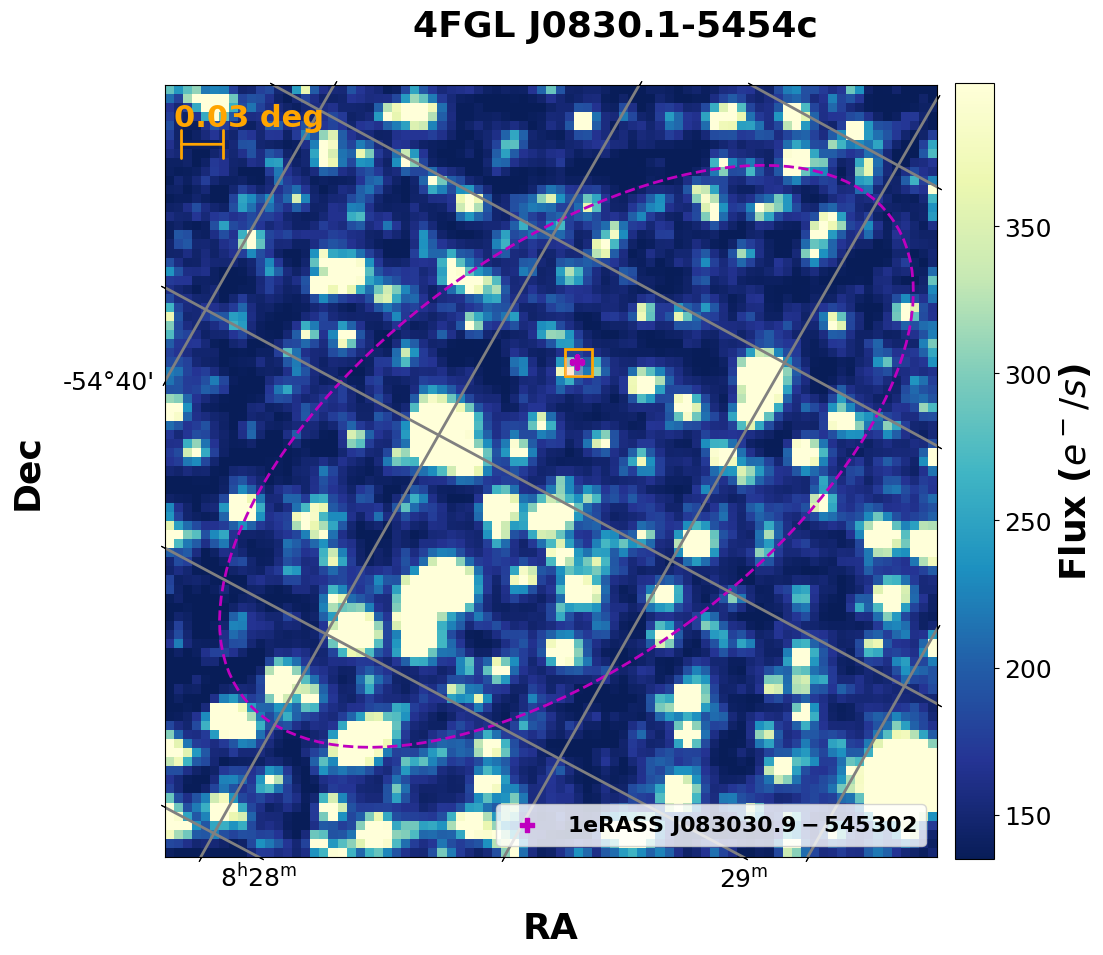} & 
        \hspace{0.01\textwidth}
        \includegraphics[width=0.4\textwidth]{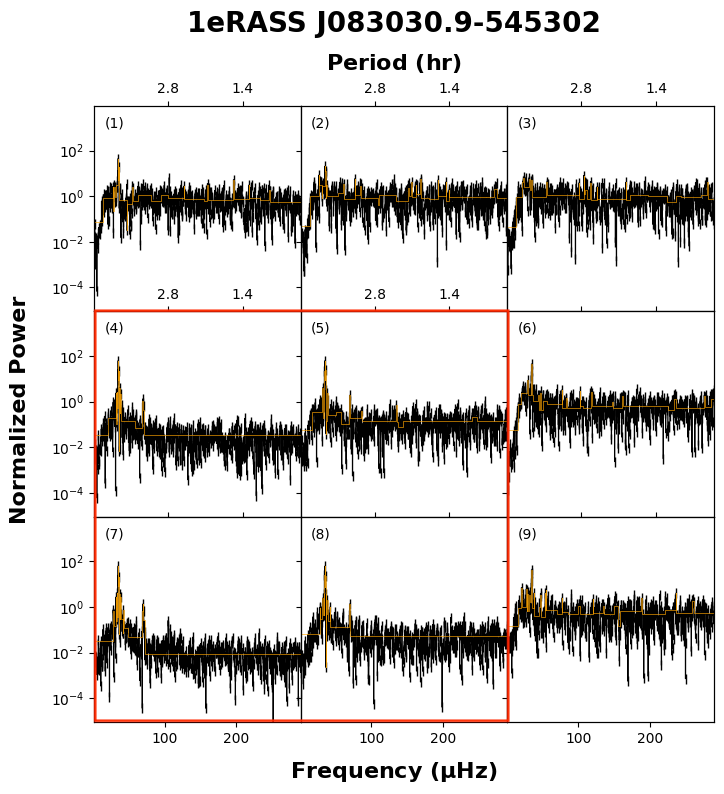}   \\
        \parbox{0.48\textwidth}{(a) TESS skymap centered on 4FGL J0830.1-5454c. Magenta plus sign marks the position of 1eRASS J083030.9-545302. Purple dashed line identifies the Fermi 95\% error ellipse. The orange box indicates the TESS pixel region shown in panel (b).} & 
        \parbox{0.4\textwidth}{(b) Pixel wise PDS of the 1eRASS J083030.9-545302 field. The pixels within the red box are used as an aperture mask for obtaining the PDS shown in panel (c).} \\[10pt]
        \includegraphics[width=0.42\textwidth]{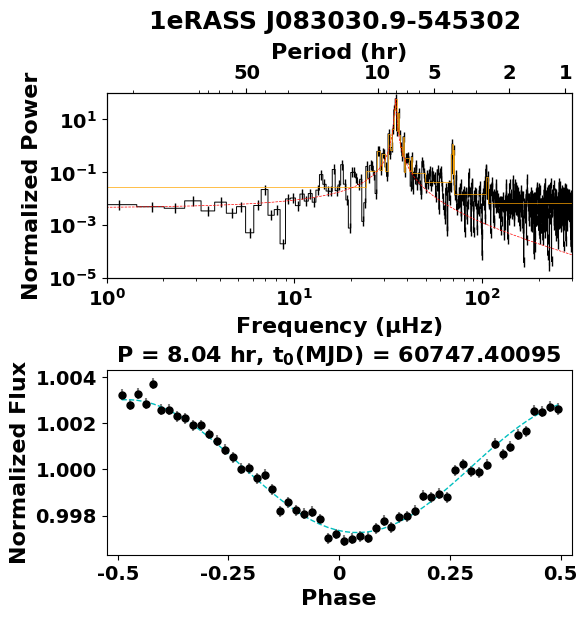} & 
        \hspace{0.01\textwidth}
        \includegraphics[width=0.45\textwidth]{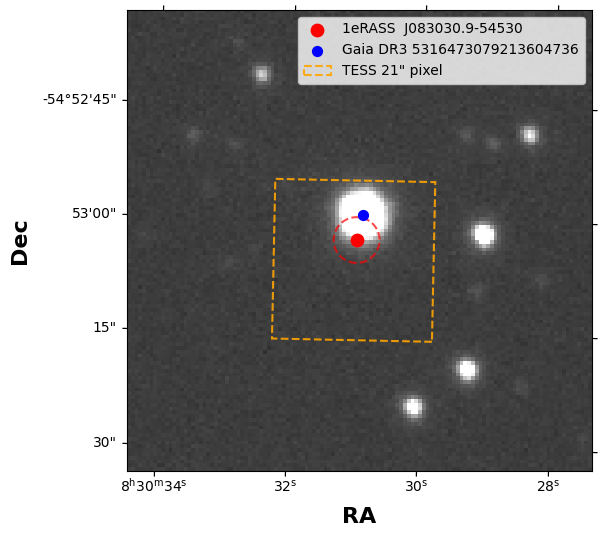}   \\
        \parbox{0.42\textwidth}{(c) PDS and folded light curve of the four pixels near 1eRASS J083030.9-545302. Light curve are folded on P = 8.04 hr.}  & 
        \parbox{0.45\textwidth}{(d) SkyMapper $r$-band image around 1eRASS 083030.9-545302. The orange dashed square marks the central pixel of the 3×3 TESS pixel region shown in panel (b), while the red dashed ellipse indicates the eROSITA positional uncertainty.}\\
    \end{tabular}
    \caption{TESS Period Search and Phase Light curves in the Region of 4FGL J0830.1-5454c and 1eRASS J083030.9-545302.}
    \label{fig:J0830}
\end{figure}

\begin{figure}[h!]
    \centering
    \begin{tabular}{cc} 
        \includegraphics[width=0.48\textwidth]{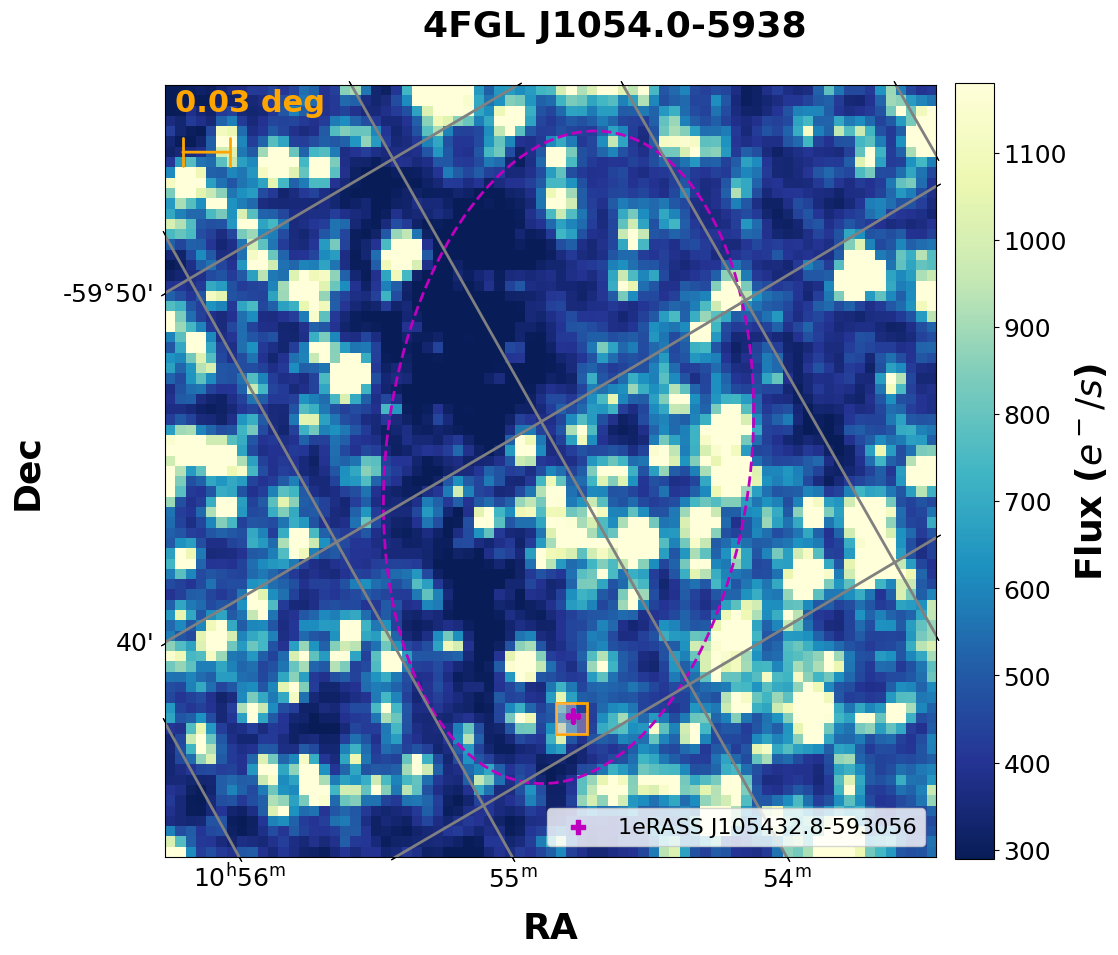} & 
        \hspace{0.01\textwidth}
        \includegraphics[width=0.4\textwidth]{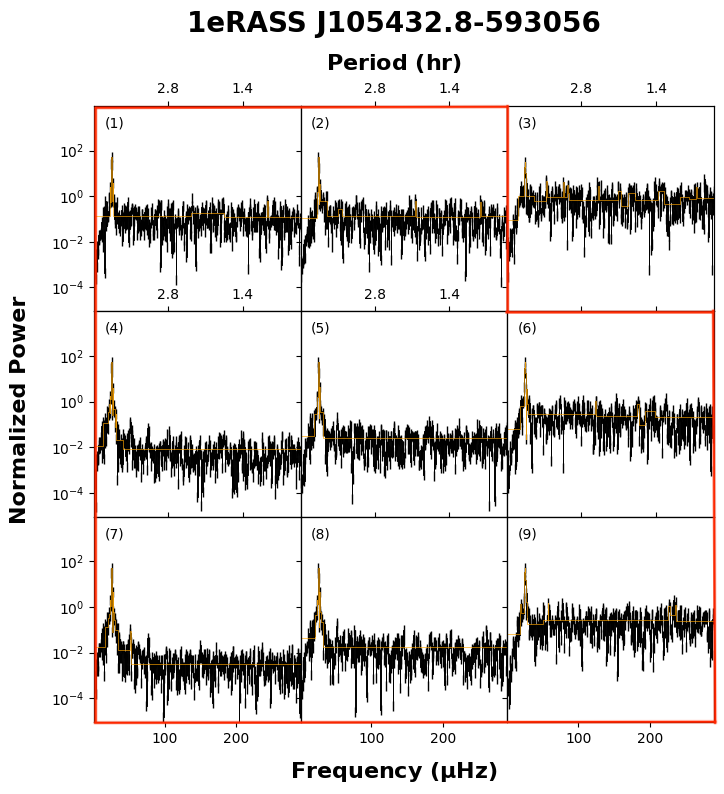}   \\
        \parbox{0.48\textwidth}{(a) TESS skymap centered on 4FGL J1054.0-5938. Magenta plus sign marks the position of 1eRASS J105432.8-593056. Purple dashed line identifies the Fermi 95\% error ellipse. The orange box indicates the TESS pixel region shown in panel (b).} & 
        \parbox{0.4\textwidth}{(b) Pixel wise PDS of the 1eRASS J105432.8-593056 field. The pixels within the red box are used as an aperture mask for obtaining the PDS shown in panel (c).} \\[10pt]
        \includegraphics[width=0.42\textwidth]{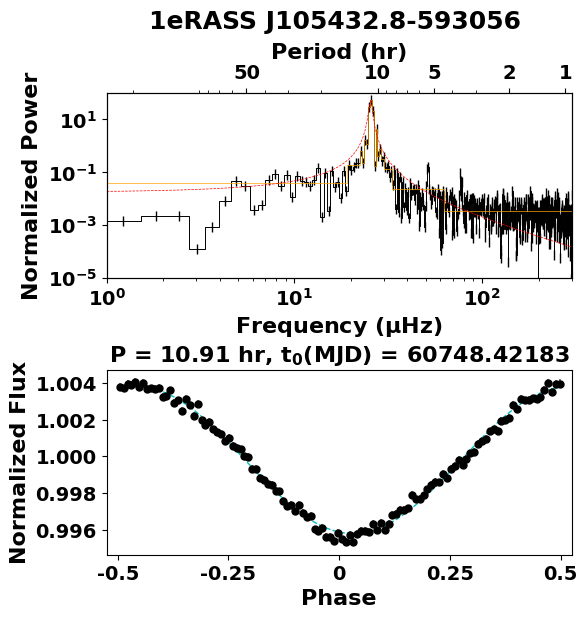} & 
        \hspace{0.01\textwidth}
        \includegraphics[width=0.45\textwidth]{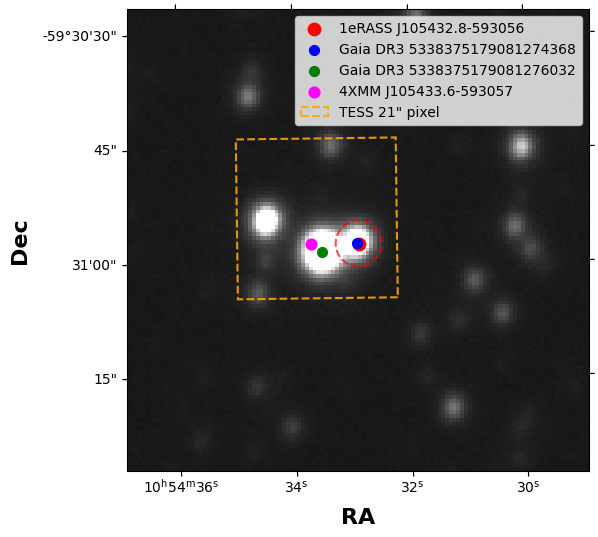}\\
        \parbox{0.42\textwidth}{(c) PDS and folded light curve the four pixels near 1eRASS J105432.8-593056. Light curve are folded on P = 10.91 hr.}  & 
        \parbox{0.45\textwidth}{(d)SkyMapper $r$-band image around 1eRASS J105432.8-593056. The orange dashed square marks the central pixel of the 3×3 TESS pixel region shown in panel (b), while the red dashed ellipse indicates the eROSITA positional uncertainty.}\\
    \end{tabular}
    \caption{TESS Period Search and Phase Light curves in the Region of 4FGL J1054.0-5938 and 1eRASS J105432.8-593056. }
    \label{fig:J1054}
\end{figure}

\begin{figure}[h!]
    \centering
    \begin{tabular}{cc} 
        \includegraphics[width=0.48\textwidth]{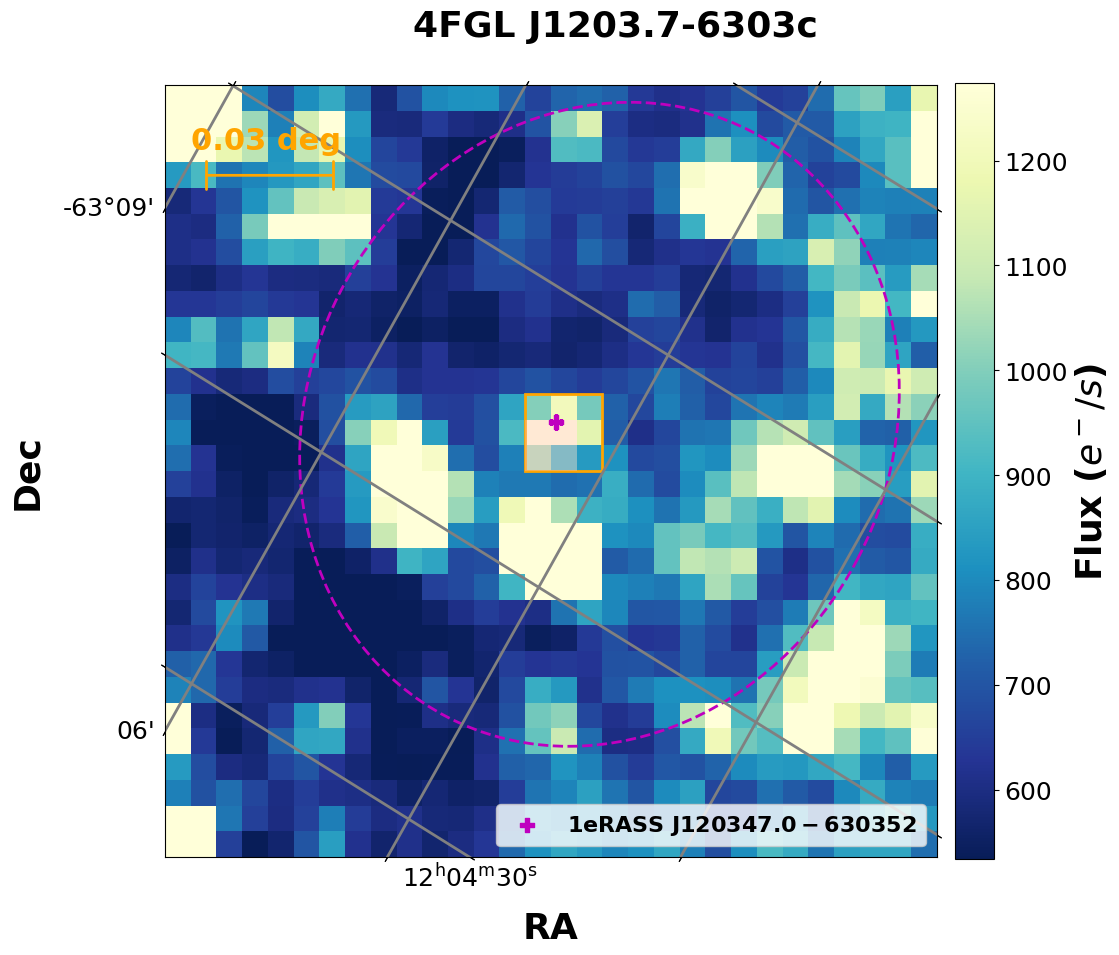} & 
        \hspace{0.01\textwidth}
        \includegraphics[width=0.4\textwidth]{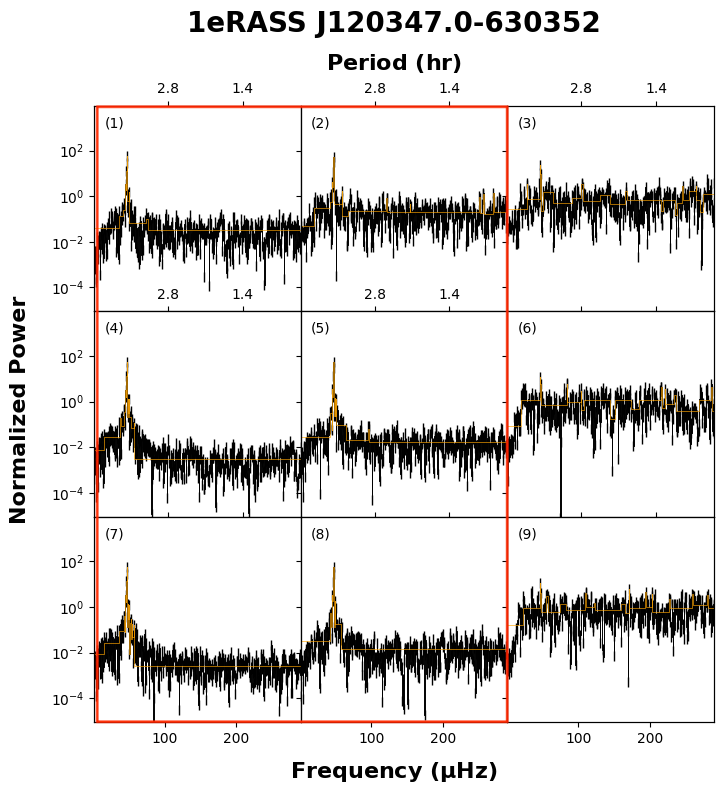}   \\
        \parbox{0.48\textwidth}{(a) TESS skymap centered on 4FGL J1203.7-6303c. Magenta plus sign marks the position of 1eRASS J120347.0-630352. Purple dashed line identifies the Fermi 95\% error ellipse. The orange box indicates the TESS pixel region shown in panel (b).} & 
        \parbox{0.4\textwidth}{(b) Pixel wise PDS of the 1eRASS J120347.0-630352  field. The pixels within the red box are used as an aperture mask for obtaining the PDS shown in panel (c).} \\[10pt]
        \includegraphics[width=0.42\textwidth]{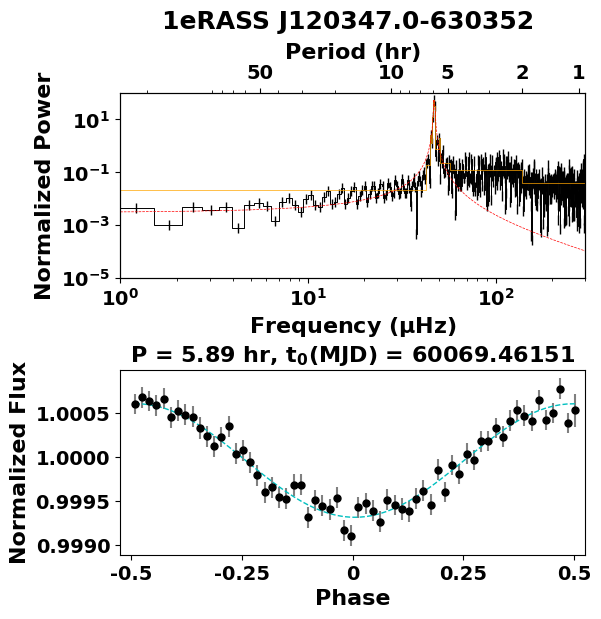}&
        \hspace{0.01\textwidth}
        \includegraphics[width=0.45\textwidth]{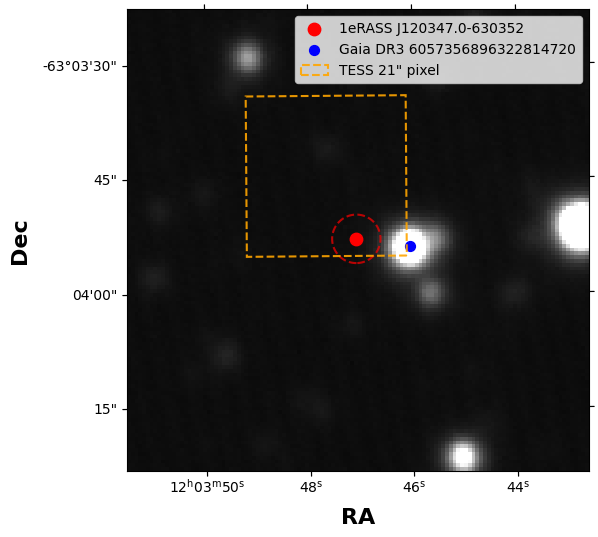}\\
        \parbox{0.42\textwidth}{(c) PDS and folded light curve of the 16 pixels coinciding 1eRASS J120347.0-630352 . Light curve are folded on P = 5.89 hr.}  & 
        \parbox{0.45\textwidth}{(d)SkyMapper $r$-band image around 1eRASS J120347.0-630352. The orange dashed square marks the central pixel of the 3×3 TESS pixel region shown in panel (b), while the red dashed ellipse indicates the eROSITA positional uncertainty.}\\
    \end{tabular}
    \caption{TESS Period Search and Phase Light curves in the Region of 4FGL J1203.7-6303c and 1eRASS J120347.0-630352.}
    \label{fig:J1203}
\end{figure}

\begin{figure}[h!]
    \centering
    \begin{tabular}{cc} 
        \includegraphics[width=0.48\textwidth]{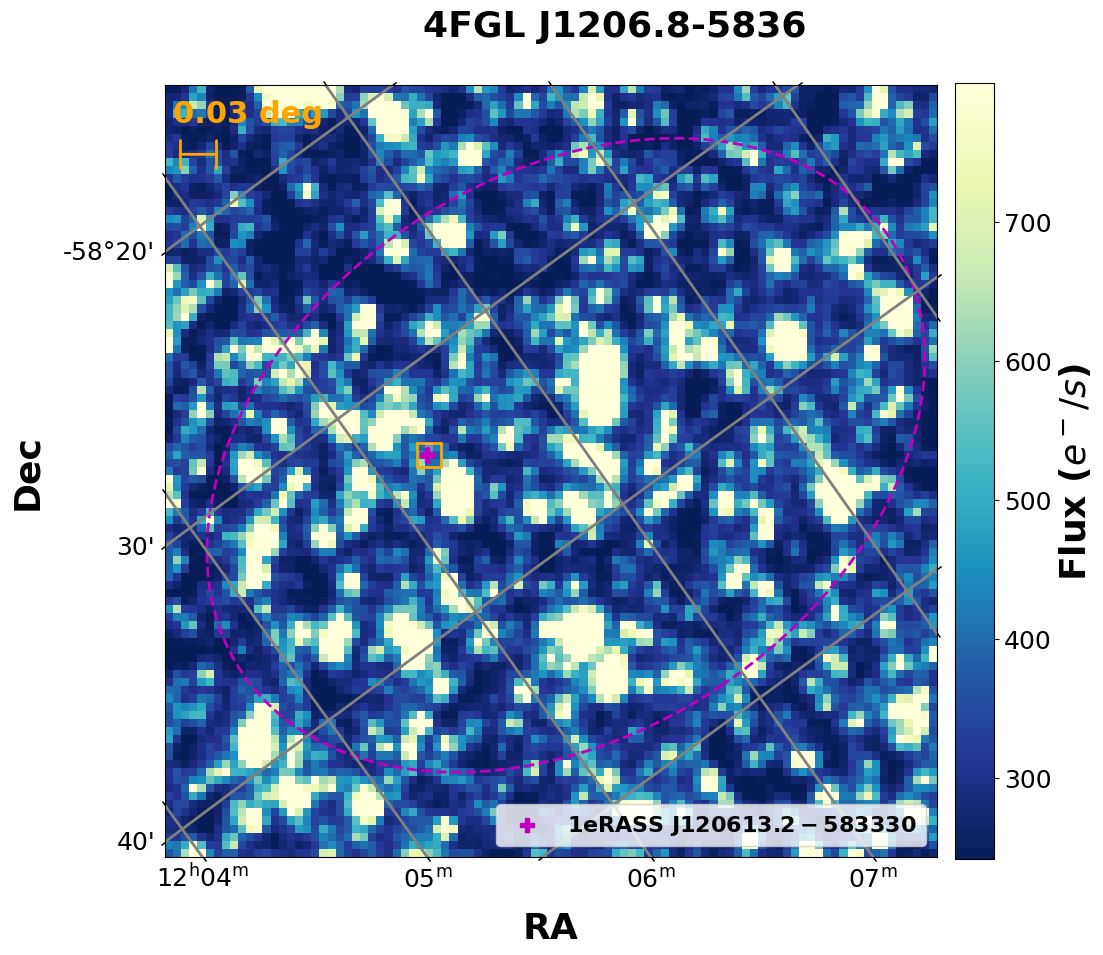} & 
        \hspace{0.01\textwidth}
        \includegraphics[width=0.4\textwidth]{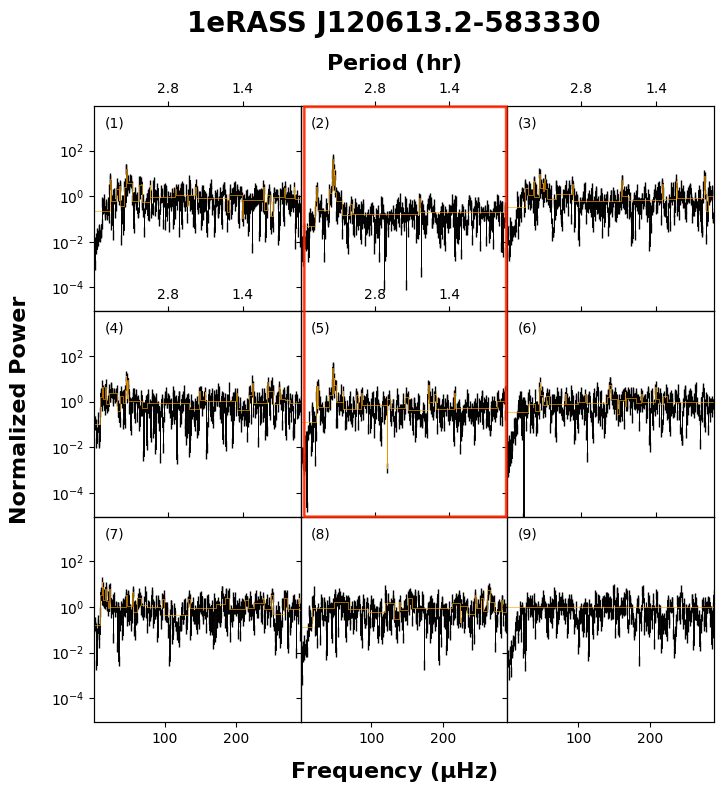}   \\
        \parbox{0.48\textwidth}{(a) TESS skymap centered on 4FGL J1206.8-5836. Magenta plus sign marks the position of 1eRASS J120613.2-583330. Purple dashed line identifies the Fermi 95\% error ellipse. The orange box indicates the TESS pixel region shown in panel (b).} & 
        \parbox{0.4\textwidth}{(b) Pixel wise PDS of the 1eRASS J120613.2-583330 field. The pixels within the red box are used as an aperture mask for obtaining the PDS shown in panel (c).} \\[10pt]
        \includegraphics[width=0.42\textwidth]{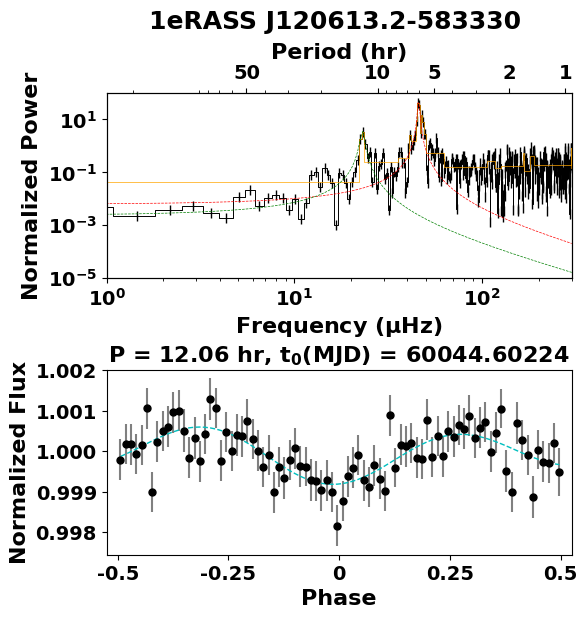}& 
        \hspace{0.01\textwidth}
        \includegraphics[width=0.45\textwidth]{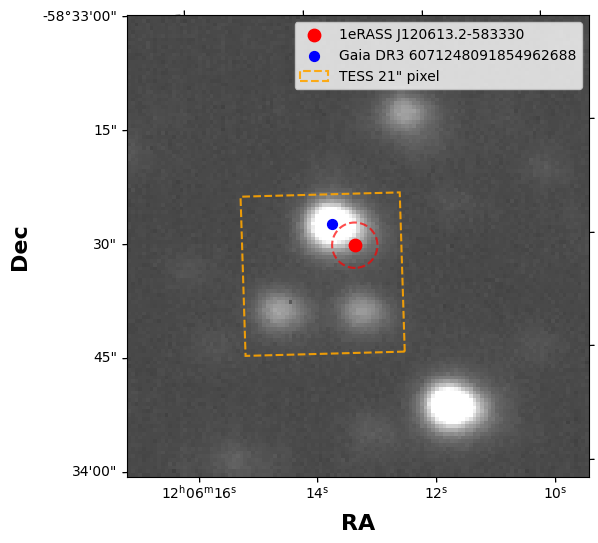}\\
        \parbox{0.42\textwidth}{(c) PDS and folded light curve of the 9 pixels coinciding 1eRASS J120613.2-583330. Light curve are folded on P = 12.06 hr.}  &         
        \parbox{0.45\textwidth}{(d)SkyMapper $r$-band image around 1eRASS J120613.2-583330. The orange dashed square marks the central pixel of the 3×3 TESS pixel region shown in panel (b), while the red dashed ellipse indicates the eROSITA positional uncertainty.}\\
    \end{tabular}
    \caption{TESS Period Search and Phase Light curves in the Region of 4FGL J1206.8-5836 and 1eRASS J120613.2-583330. }
    \label{fig:J1206}
\end{figure}

We conducted data analysis on TESS sky region data centered on the coordinates of the 183 eROSITA sources, each covering an area of 3 pixels ${\times}$ 3 pixels (63$^{\prime\prime}$ ${\times}$ 63$^{\prime\prime}$). For comparison, the typical positional uncertainty of an eROSITA source is less than 10$^{\prime\prime}$, smaller than the TESS pixel scale of 21$^{\prime\prime}$. Therefore, additional cross-matching with Gaia is required. The positional relationships among these detections are illustrated in Figs.~\ref{fig:J0830}-\ref{fig:J1206}(d), which show the SkyMapper $r$-band images of the fields\citep{2019PASA...36...33O}. To search for optical periodicity, power spectra were computed for each pixel in the 3 ${\times}$ 3 region around the eROSITA source. The presence of peaks in the power density spectra (PDS) of individual pixels indicates the existence of periodic signals. We selected these sources with periodic signals in the optical band and conducted further analysis on them. Folded light curves are plotted using the periods obtained from pixels showing periodic signals. 

TESS archival data are obtained from TESSCUT \citep{2019ascl.soft05007B} and extracted from the Full Frame Images (FFI). All data were processed with Lightkurve \citep{2018ascl.soft12013L}. Light curve files are created using the \verb|to_lightcurve()| function from calibrated target-pixel data. For TESSCUT data, the aperture mask is manually selected on the basis of the presence of peak profiles in the PDS of individual pixels. \verb|remove_nans()| is used to remove the infinite or NaN values from the light curves. Outliers exceeding the 5-$\sigma$ level in the light curves are clipped with \verb|remove_outliers()|. To identify periodicities in the optical flux, Power Density Spectra are generated from cleaned, unbinned light curves using the \verb|to_periodogram()| function. 

Power Density Spectra are calculated using the \texttt{Lomb-Scargle} method and normalized to \texttt{power spectral density} \citep{balona20}. Significant peak profiles in power density spectra are determined with \textit{Bayesian block} analysis \citep{bayes} with 95\% statistical significance using \texttt{Astropy}. The peak profiles obtained from \textit{Bayesian block} analysis are fitted with a \textit{Lorentzian} profile \citep{belloni02}. \texttt{Scipy} is used to estimate the significance of the peak profiles. From curve fitting, Q-value ($\frac{\nu}{\Delta\nu}$) \citep{casella05}, RMS amplitude \footnote{\url{https://heasarc.gsfc.nasa.gov/docs/xte/recipes/pca\_fourier.html}} [see Eq.~\ref{eq_rms}] are calculated and listed in Table~\ref{tab:details}. 

\begin{align}
RMS = & 100 \times \sqrt{\frac{A}{\overline{Flux}}} \%, \label{eq_rms} \\
where, ~ A = & \frac{\pi}{2} \times Normalization \times FWHM, \nonumber \\
 = & ~ Flux ~ under ~ Lorentzian ~ function; \nonumber \\
Normalization = & ~ Power ~ at ~ peak ~ frequency, \nonumber \\
FWHM = &~ Full ~ width ~ half ~ maxima. \nonumber  
\end{align}

The TESS skymap is plotted for all sources, with sizes varying according to the Fermi 95\% error ellipse. The Fermi 95\% error ellipses are plotted with purple dashed ellipses. The positions of the eROSITA sources are marked on the skymap with a magenta plus sign. The red box centered on the eROSITA source represents the 3 ${\times}$ 3 pixel sky region, which is taken for further analysis. All skymaps are shown in Figs.~\ref{fig:J0830}-\ref{fig:J1206}(a). In Figs.~\ref{fig:J0830}-\ref{fig:J1206}(b) pixel-wise PDS for the 3 ${\times}$ 3 pixel orange box is shown. The orange curve shows the Bayesian blocks with 95\% statistical significance. Significant peak profiles are fitted with the Lorentzian model and the pixels showing significant peaks are taken as aperture mask, indicated by a red square. Next, integrating those pixels with significant peak profiles helps us to better characterize the period peaks, as shown in red and green color. The results of the analysis are shown in Table~\ref{tab:details}. 

To reduce the risk of mis-association, we searched for possible Gaia counterparts in the Gaia Archive. Since stellar proper motions may cause noticeable shifts between the Gaia DR3 reference epoch (J2016.0) and the eROSITA observations (2020), we propagated the Gaia coordinates to the eROSITA epoch using the catalogued proper motions. The propagated positions were then cross-matched with the eROSITA error ellipses to identify possible counterparts.

\section{Results} \label{sec:result}
We searched 183 sources and identified 21 showing periodic signals, among which four exhibited sinusoidal modulation in their folded light curves. All 4 spider candidates show periodicities ranging from 3 to 13 hours, more information about them is given in Table~\ref{tab:details}. TESS R.A. and Decl. are the average Right Ascension and Declination of TESS pixels showing periodicity. $T_{mag}$ is TESS magnitude calculated from the average flux.  The reported periods correspond to the highest-power peak identified in the PDS. In cases where both a fundamental and its first harmonic were visible, we adopted the fundamental if folding the light curve at twice the candidate frequency produced two similar orbits, and adopted the harmonic otherwise. Sky region gives the size of the TESS cutout downloaded around the 4FGL source. Aperture mask indicates the set of pixels used for optical periodicity analysis, highlighted in red box in the PDS images. $Amp$ is the peak-to-peak amplitude of the folded light curve and $Rel. Amp.$ is the relative amplitude, defined as $\mathrm{Rel.\,Amp.} = \mathrm{Amp}/\langle F \rangle$, where $\langle F \rangle$ is the median flux of the light curve.

\begin{table}[H]
\raggedright
\caption[]{Details of Candidates \label{tab:details}}
\setlength{\tabcolsep}{1pt} 
\scriptsize
\begin{tabular}{ccccccccccccccc}
  \hline\noalign{\smallskip}
eROSITA name & R.A. & Decl. & TESS & Gaia R.A. & Gaia Decl. & Period & T$_{mag}$  & Sky region & Mask & Peak freq & $Amp$ & $Rel. Amp.$ & Q-value & RMS\\
(1eRASS) & ($^{\circ}$) & ($^{\circ}$) & sector & ($^{\circ}$) & ($^{\circ}$) & (hr) & & ($^{\circ} \times^{\circ}$) & pixel(s) & ($\mu$Hz) & (e$^{-} s^{-1}$) & (\%) &  & (\%)\\
\hline\noalign{\smallskip}
J083030.9-545302 & 127.6291 & -54.8841 & 90 & 127.6225 & -54.8838 & 8.04 & 14.3 & 0.5 $\times$ 0.5 & 4 & 69.224 & 2.23 & 0.76 & 149.20 & 0.21  \\
J105432.8-593056 & 163.6370 & -59.5158 & 90 & 163.6370 & -59.5155 & 10.91 & 13.5 & 0.44 $\times$ 0.44 & 8 & 51.505 & 5.17 & 0.86 & 59.37 & 0.26\\
J120347.0-630352 & 180.9462 & -63.0647 & 65 & 180.9518 & -63.0646 & 5.89 & 13.1 & 0.18 $\times$ 0.18 & 6 & 47.105 & 7.6 & 0.65 & 179.89 & 0.05\\
J120613.2-583330 & 181.5552 & -58.5586 & 64 & 181.5607 & -58.5573 & 12.06 & 13.9 & 0.55 $\times$ 0.55 & 2 & 45.905 & 1.17 & 0.31 & 102.69 & 0.09\\
  \hline
\end{tabular}
\end{table}

We modeled the TESS light curves using sinusoidal-like fitting to characterize the optical variability. For sources exhibiting simple periodic modulations, we employed a single-sinusoid model:
\begin{align}
f(t) = a + b \sin(\omega t + \phi),
\end{align}
where \(\omega = \frac{2\pi}{P}\) is the angular frequency corresponding to the best-fit period.
For sources exhibiting asymmetric double-peaked modulations—characterized by unequal minima depths or visible harmonic structures in the power density spectrum (PDS)—we used a two-harmonic model:
\begin{align}
f(t) = a + b_1 \sin(\omega t + \phi_1) + b_2 \sin(2 \omega t + \phi_2).
\end{align}
These trigonometric models provide a convenient approximation of the physical variations in the system.

\section{Discussion} \label{sec:discussion}
\subsection{\texorpdfstring{$\gamma$-ray properties}{gamma-ray properties}}
To better understand the nature of these systems, we also examined the classifications of their associated $\gamma$-ray sources. The $\gamma$-ray parameters of the four candidates, including their spectral models, energy fluxes, and variability indices, are summarized in Table~\ref{tab:GammaRay}. Among our four candidates, one source,  4FGL J0830.1-5454c, is included in the machine-learning-based catalog by \citet{2024A&A...684A.208M}, which assigns pulsar probabilities ($P_{\mathrm{PSR}}$) to all unassociated \textit{Fermi}-LAT sources. In their results, 909 sources are identified as pulsar candidates with $P_{\mathrm{PSR}} > 0.02$. For 4FGL J0830.1-5454c, the assigned $P_{\mathrm{PSR}}$ is 0.06. While the value is not particularly high, it places the source among the higher-ranked pulsar candidates in the catalog. For the non-candidate sources in our sample, the $\gamma$-ray spectra and variability indices suggest that most are pulsar-like rather than blazar-like. Several of them are also included in the machine-learning-based catalog, but no significant optical modulation was detected in their TESS light curves. The remaining sources, including the other three candidates, are not listed in \citet{2024A&A...684A.208M}, likely because their association probabilities or spectral significances fall below the thresholds adopted in that catalog.

\subsection{Multi-wavelength associations}
The calculated G-band, X-ray and $\gamma$-ray luminosities of the four candidates are listed in Table \ref{tab:luminosity}, providing a basis for multiwavelength comparison with known redback systems. The X-ray and $\gamma$-ray data for known redbacks were taken from \cite{2025ApJ...994....8K}, while the Gaia G-band luminosities were calculated using the distances and the magnitude from the Gaia DR3 catalog.

\begin{figure}
    \centering
    \includegraphics[width=0.45\linewidth]{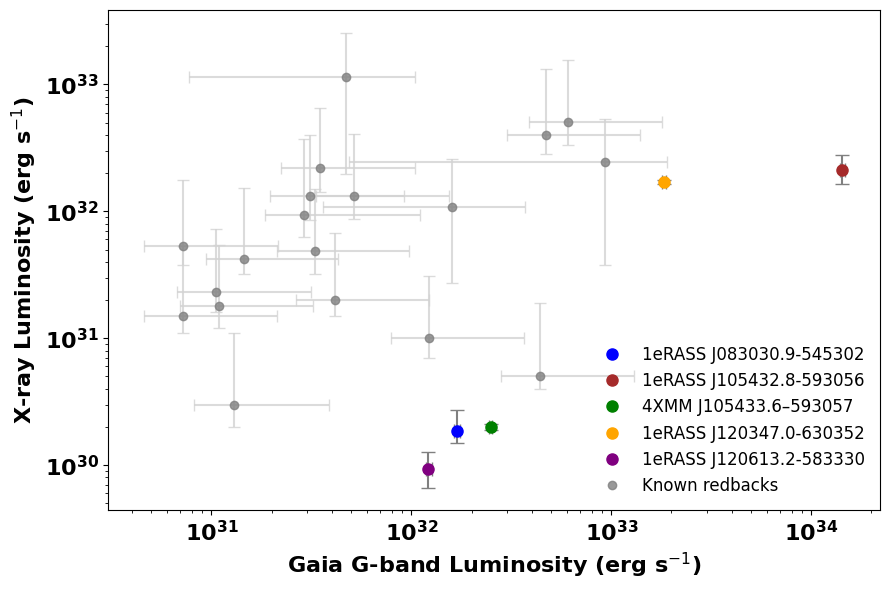}
    \caption{X-ray versus optical luminosity. Known redbacks from \cite{2025ApJ...994....8K} are shown for comparison.}
    \label{fig:x-ray vs optical}
\end{figure}

\begin{figure}
    \centering
    \includegraphics[width=0.45\linewidth]{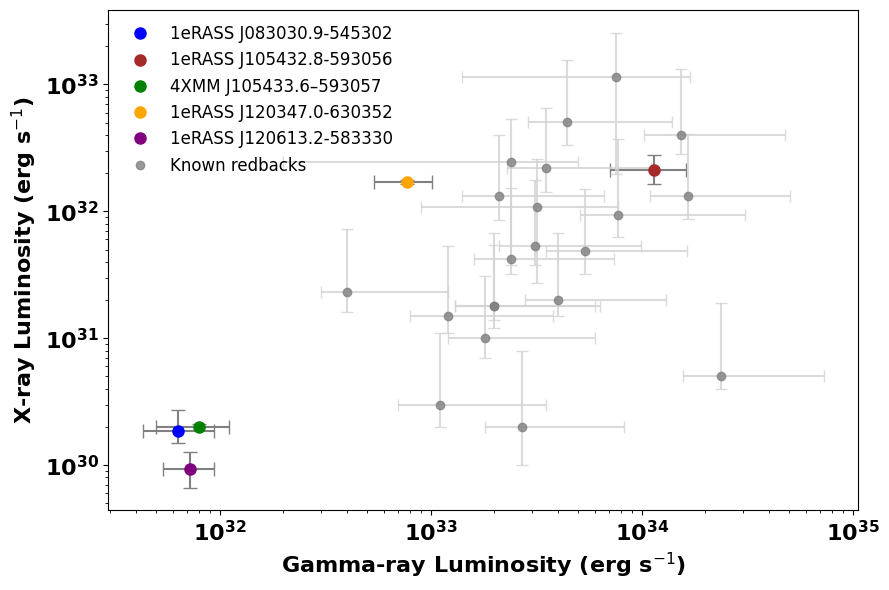}
    \caption{X-ray versus $\gamma$-ray luminosity for the same candidates, compared with known redbacks from \cite{2025ApJ...994....8K}.}
    \label{fig:x-ray vs Gamma}
\end{figure}

In Fig. \ref{fig:x-ray vs optical} and Fig. \ref{fig:x-ray vs Gamma}, the remaining candidates occupy a region in the luminosity plane that partially overlaps with the known redback population. The optical brightness of the counterparts matches expectations for irradiated sub-solar companions. The approximate correlation between X-ray and $\gamma$-ray luminosities may partly reflect a common dependence on the pulsar spin-down power ($\dot{E}$), as systems with higher $\dot{E}$ tend to exhibit stronger emission across multiple energy bands. A clearer dependence could be tested in future by incorporating $\dot{E}$ information once pulsation measurements become available. Although the scatter is larger than for confirmed systems—partly due to distance uncertainties—the consistency in multiwavelength luminosity scaling supports their possible nature as MSP binaries. 

Based on these multi-wavelength characteristics, we discuss the four candidates individually below. We note that three of our four candidates are located at low Galactic latitude ($|b| < 5^\circ$), where extinction and diffuse backgrounds increase the chance of source confusion. However, this does not affect the identification as potential spider candidates. Their overall properties such as positional consistency with $\gamma$-ray and X-ray sources, and light curve modulations, still support their candidacy.

\begin{table}[]
    \centering
    \begin{tabular}{ccccccc}
    \hline
     4FGL name & R.A. & Decl. & Spectrum Type & 0.1-100~GeV Flux & Variability Index & Significance\\
     (4FGL) & ($^{\circ}$) & ($^{\circ}$) & & ($\times 10^{-12}erg/cm^2/s$) & & ($\sigma$) \\
     \hline
     J0830.1-5454c & 127.53 & -54.91 & log-parabola & $2.69 \pm 0.66$ & $9.41$ & 6.05 \\
     J1054.0-5938  & 163.51 & -59.64 & log-parabola & $4.33 \pm 1.59$ & $9.33$ & 5.15 \\
     J1203.7-6303c & 180.94 & -63.06 & log-parabola &  $ 8.38 \pm 2.36$ & $17.49$ & 6.64\\
     J1206.8-5836 & 181.71 & -58.62 & log-parabola & $4.04 \pm 0.96$ & $9.81$ & 5.55\\ 
     \hline
    \end{tabular}
    \caption{$\gamma$-ray properties of the four candidate spider systems. All values are taken from the 4FGL-DR4 catalog.}
    \label{tab:GammaRay}
\end{table}

\subsection{1eRASS J083030.9-545302}
For 1eRASS J083030.9-545302, no Gaia sources are found within the 1-$\sigma$ positional uncertainty. The nearest Gaia source Gaia DR3 5316473079213604736 is found with a separation of 3.465". The photogeometric distance is about 375 pc, leading to an X-ray luminosity of $L_{X} = 1.85^{+0.86}_{-0.34} \times 10^{30} erg\, s^{-1}$. This luminosity is lower than the typical RB range ($10^{31}$–$10^{32} erg\,s^{-1}$), and similar to the BW systems \citep{2018ApJ...864...23L,2022MNRAS.511.5964Z}. 

\subsection{1eRASS J105432.8-593056 / 4XMM J105433.6–593057}
For 1eRASS J105432.8-593056, we identified two Gaia sources located close to the X-ray positions. Gaia DR3 5338375179081274368 is located near the eROSITA source 1eRASS J105432.8-593056 and has a mean G-band magnitude of 14.35. Beyond the eROSITA detection, we conducted a search for potential XMM-Newton counterparts \citep{2020A&A...641A.136W} and found one associated source, 4XMM J105433.6–593057\footnote{\url{http://xmm-catalog.irap.omp.eu/source/208402104010002}}, which lies close to Gaia DR3 5338375179081276032 and is located $6.3^{\prime\prime}$ from the eROSITA position. Their relative positions are illustrated in the SkyMapper $r$-band image (Fig. \ref{fig:J1206}(d)). This configuration reveals two potential X-ray–optical counterpart combinations. 

The eROSITA-Gaia combination yields an X-ray luminosity of $L_{X} = 2.12^{+0.66}_{-0.48} \times 10^{32} erg\,s^{-1}$, consistent with typical redback systems. The XMM-Gaia combination gives $L_{X} = 2.00^{+0.10}_{-0.10} \times 10^{30} erg\,s^{-1}$. The XMM source is detected with high significance, has a 0.2-12 keV flux of $(1.09 \pm 0.04) \times 10^{-13} erg\,cm^{-2}\,s^{-1}$, and its hardness ratios (HR1 = $0.54 \pm 0.02$; HR2 = $-0.16 \pm 0.02$) indicate a moderately soft spectrum.

Either combination could contribute to the observed TESS optical modulation, although their true associations remain uncertain. For both possible X-ray–Gaia combinations, we compute the corresponding luminosities and list them in Table~\ref{tab:luminosity}. Their positions in the luminosity–luminosity diagrams are also indicated in Figs.~\ref{fig:x-ray vs optical} and \ref{fig:x-ray vs Gamma}.

\subsection{1eRASS J120347.0-630352}
No Gaia sources are located within the 1-$\sigma$ positional uncertainty of the eROSITA source. A bright Gaia source (DR3 6057356896322814720) lies outside this region, at a separation of 7.08", and is considered as a possible optical counterpart. Other fainter Gaia sources are unlikely to contribute to the TESS detection. The observed TESS modulation may originate from this Gaia source, which could be distinct from the X-ray emitter. Assuming a distance of 877 pc, the derived X-ray luminosity is $L_{X} = 1.7^{+0.07}_{-0.05} \times 10^{32} erg\,s^{-1}$, within the typical range for redback systems.

\subsection{1eRASS J120613.2-583330}
The folded light curve shown in Fig. \ref{fig:J1206}(c) displays a double-peaked modulation that is accurately modeled using a two-component harmonic sine function. The unequal peak structure, especially the deeper minimum, reflects an intrinsic asymmetry in the optical modulation of the system. Such double-humped light curves are commonly observed in compact binary systems where ellipsoidal modulation, irradiation effects, or Doppler beaming can dominate the optical variability. The relatively broad and deep minima may also indicate partial eclipses or a heated face scenario, where the companion is irradiated by a compact object \citep{2019A&A...621L...9Y}.

For 1eRASS J120613.2-583330, there is one source named Gaia DR3 6071248057488411648 located in its positional error ellipse, with the angular separation of 1.385". But with a mean g-band magnitude of 18.76, it's too faint and not consistent with our searching in TESS. Expanding the Gaia matching area, Gaia DR3 6071248091854962688 with a mean g-band magnitude of 14.1 and photometric variability flag is marked as "variable". But with the distance of about 386 pc, the X-ray luminosity is estimated to be $L_{X} = 9.31^{+3.48}_{-2.65} \times 10^{29} erg\,s^{-1}$. 

Given the relatively low X-ray luminosity and the TESS light curve shown in Fig. \ref{fig:J1206}(c), the modulation exhibits a tentative M-shaped trend, which could be indicative of a binary system of two main-sequence stars (EW-type), though this interpretation remains uncertain. Such systems typically exhibit nearly sinusoidal variations with alternating minima of comparable depths \citep{2017RAA....17...87Q, 2025NatSR..1528369S}, which is broadly consistent with the observed variability.

\begin{table}[ht]
\centering
\begin{minipage}[]{\linewidth}
\caption[]{Luminosity correlation of the candidates. $L_X$ shows 0.2–2.3 keV energy band X-ray luminosity. Here flux unit is $10^{-14}$ erg s$^{-1}$cm$^{-2}$, and lum unit is $10^{30}$erg s$^{-1}$.  \label{tab:luminosity}}
\end{minipage}
\setlength{\tabcolsep}{1pt} 
\scriptsize
\begin{tabular}{lcccccccccc}
  \hline\noalign{\smallskip}
  4FGL name & X-ray name &Gaia ID&  $G_{\rm mean}$ & $L_{\rm G}$ & $F_{X}$ & $L_{X}$& $L_{\gamma}$ & Gaia-X & X-TESS & Gaia-TESS\\
   (4FGL) & & (DR3) &  & (lum unit) & (flux unit) & (lum unit) & (lum unit) & ($^{\prime\prime}$) & ($^{\prime\prime}$) & ($^{\prime\prime}$)\\
  \hline\noalign{\smallskip}
    J0830.1-5454c & \makecell{1eRASS\\J083030.9-545302} & 5316473079213604736 & 13.67 & $169^{+6.4}_{-5.4} $ & $11^{+4.5}_{-1.7} $ & $1.85^{+0.86}_{-0.34} $ &$63.4^{+21.4}_{-20.4}$ & 3.476 & 2.595 & 6.067\\
    J1054.0-5938$^{a}$  & \makecell{1eRASS\\J105432.8-593056} & 5338375179081274368 & 14.35 & $14200^{+542}_{-228}$ & $8.05^{+2.1}_{-1.7}$ & $212^{+66.0}_{-47.8}$ & $11400^{+4780}_{-4300}$ & 0.707 & 6.049 & 6.180\\
    J1054.0-5938$^{b}$ & \makecell{4XMM \\ J105433.6–593057} & 5338375179081276032 & 13.35 & $249^{+2.9}_{-3.0}$ & $10.86^{+0.42}_{-0.42}$ & $2.00^{+0.10}_{-0.10}$ & $79.7^{+30.4}_{-29.8}$ & 1.787 & 3.446 & 4.477\\
    J1203.7-6303c & \makecell{1eRASS\\J120347.0-630352} & 6057356896322814720 & 12.93 & $1840^{+47.1}_{-49.8}$ & $3.64^{+1.32}_{-1.10}$ & $170^{+7.31}_{-5.38}$ &$772^{+242}_{-232}$ & 7.088 & 9.138 & 14.315\\
    J1206.8-5836 & \makecell{1eRASS\\J120613.2-583330} & 6071248091854962688  & 14.1 & $121^{+5.5}_{-3.0} $ & $5.21^{+1.6}_{-1.4}$ &$0.93^{+0.35}_{-0.27} $ &$72.1^{+21.1}_{-18.5}$ & 4.177 & 5.736 & 6.707\\
  \hline
\end{tabular}
\\[5pt]
\begin{minipage}{\linewidth}
\footnotesize
\raggedright
\textbf{Notes.} J1054.0-5938$^{a}$ and J1054.0-5938$^{b}$ correspond to two X-ray–Gaia combinations: 1eRASS J105432.8-593056 with Gaia DR3 5338375179081274368 (eROSITA) and 4XMM J105433.6–593057 with Gaia DR3 5338375179081276032 (XMM-Newton).

The last three columns (Gaia-X, X-TESS, Gaia-TESS) give the angular separations between Gaia and eROSITA, eROSITA and TESS centroid, and Gaia and TESS centroid, respectively. 
\end{minipage}
\end{table}

\section{Conclusion} \label{sec:conclusion}
We search for new spider candidates through cross-matching in multiple wavelengths, including $\gamma$-ray, X-ray, and optical data. We analyze TESS data for 183 eROSITA candidates, selecting those with sinusoidal optical light curves. Our search identifies four candidates, corresponding to the following pairs: 4FGL J0830.1-5454c/1eRASS J083030.9-545302, 4FGL J1054.0-59384FGL J1054.0-5938 (with two possible X-ray associations: 1eRASS J105432.8-593056 or 4XMM J105433.6-593057), 4FGL J1203.7-6303c/1eRASS J120347.0-630352, 4FGL J1206.8-5836/1eRASS J120613.2-583330. All four candidates exhibit periodic variability on timescales of 5 to 13 hours. We analyzed the optical light curves of all four candidates and found that each exhibits variability resembling sinusoidal modulation to some extent. 
Such variability could arise from ellipsoidal distortion or irradiation effects if a pulsar is present, but other mechanisms such as stellar rotation \citep{2003MNRAS.345.1145D}, star spots \citep{2011ApJ...736..123M}, or tidal interactions in close binaries \citep{1997AJ....113.1841M} may also contribute.

We emphasize that the associations between the TESS sources and the X-ray counterparts are not fully confirmed, and in one case (4FGL J1054.0-5938), more than one plausible X-ray–optical pairing exists. Therefore, while these candidates may represent relatively X-ray faint spider systems, it is also possible that some of the observed optical variability originates from unrelated variable stars, and the systems may not host pulsars. Notably, all four candidates exhibit sinusoidal-like optical modulations, lacking the broad peaks with sharp minima typical of BW systems powered by irradiation. This feature leads us to propose that these candidates may represent RB systems. Our findings provide a new sample of spider system candidates, which will be essential for future multi-wavelength observations. Spectroscopic follow-up will be crucial to determine the nature of the companion stars, while X-ray and radio observations can confirm the presence of MSPs.

\newpage 

\clearpage

\begin{acknowledgements}
We acknowledge the use of the \textit{Fermi}-LAT Fourth Source Catalog (4FGL), provided by the Fermi Large Area Telescope collaboration, and the eROSITA source catalog, obtained from the primary instrument aboard the SRG mission, a joint Russian-German science mission.

This work also makes use of data from the European Space Agency (ESA) mission Gaia (\url{https://www.cosmos.esa.int/gaia}), processed by the Gaia Data Processing and Analysis Consortium (DPAC, \url{https://www.cosmos.esa.int/web/gaia/dpac/consortium}). Funding for DPAC has been provided by national institutions, particularly those participating in the Gaia Multilateral Agreement.

Additionally, this paper includes data collected with the TESS mission, obtained from the MAST data archive at the Space Telescope Science Institute (STScI). Funding for the TESS mission is provided by the NASA Explorer Program. STScI is operated by the Association of Universities for Research in Astronomy, Inc., under NASA contract NAS 5–26555 \citep{MAST2021}.
We also acknowledge the use of the TESSCUT tool~\citep{2019ascl.soft05007B} for extracting Full Frame Images and the Lightkurve package~\citep{2018ascl.soft12013L} for data processing.

We also acknowledge the use of SkyMapper $r$-band images. The national facility capability for SkyMapper has been funded through ARC LIEF grant LE130100104 and the survey data were processed and provided by the SkyMapper Team at ANU. SkyMapper is operated by the Australian National University's Research School of Astronomy and Astrophysics, with support from ASVO, AAL, and the Australian Government through NCRIS and NeCTAR.

XQL, PHT, and WJH thank the support from the National Natural Science Foundation of China (NSFC) under grant No. 12273122, National Astronomical Data Center, the Greater Bay Area, under grant No. 2024B1212080003, and science research grant from the China Manned Space Project under CMS-CSST-2025-A13. 
P.~S.~Pal is supported by a General Research Fund (GRF) grant from the Research Grants Council of the Hong Kong Special Administrative Region, China, HKU-RMGS Funds (207300301,207301032; PI: Prof.~Q.~A.~Parker). R.~Diwan is supported by HKU-RMGS Funds (207300301,207301033; PI: Prof.~Q.~A.~Parker). 
\end{acknowledgements}

\bibliographystyle{raa}
\bibliography{bibtex} 

\appendix 

\section{Machine-readable table for all candidates analysed.}
\begin{footnotesize}

\begin{longtable}{ccccccccc}
\caption[]{Details of Candidates analysed. The brightest probable TESS counterparts are obtained from the TIC catalog within 30 arcsec. The TESS variability is determined from flux changes in 9-pixels region around the eROSITA source position, rather than from the TIC source. \label{mrt}} \\\hline\hline

Name & eRosita Name & Ra & Dec & radec\_err & TIC-ID & TESS & Distance & TESS\\
(4FGL) & (1eRASS) & (deg) & (deg) & (arcsec) & \# & mag & (arcsec) & variability \\\hline
\endfirsthead

\caption[]{Details of Candidates analysed. The brightest probable TESS counterparts are obtained from the TIC catalog within 30 arcsec. The TESS variability is determined from flux changes in 9-pixels region around the eROSITA source position, rather than from the TIC source (continued).} \\\hline\hline

Name & eRosita Name & Ra & Dec & radec\_err & TIC-ID & TESS & Distance & TESS\\
(4FGL) & (1eRASS) & (deg) & (deg) & (arcsec) & \# & mag & (arcsec) & variability \\\hline
\endhead

\hline
\multicolumn{9}{r}{\textit{Continued on next page}} \\
\endfoot

\hline\hline
\endlastfoot

J0110.0-5246 & J010928.7-524559 & 17.370 & -52.766 & 5.180 & 616155999 & 19.06 & 2.20 & NO\\
J0137.3-3239 & J013724.5-324037 & 24.352 & -32.677 & 4.721 & 632119596 & 18.52 & 18.08 & NO\\
J0139.5-2228 & J013933.6-222746 & 24.890 & -22.463 & 5.587 & 632246878 & 19.24 & 2.15 & NO\\
J0213.2-8333 & J021225.1-833324 & 33.105 & -83.557 & 5.727 & 630694769 & 18.29 & 9.33 & NO\\
J0243.1-4218 & J024305.0-422028 & 40.771 & -42.341 & 3.362 & 651217496 & 19.09 & 0.69 & NO\\
J0251.1-1830 & J025111.5-183114 & 42.798 & -18.521 & 1.468 & 651805436 & 18.98 & 1.67 & NO\\
J0301.6-5617 & J030156.7-561507 & 45.487 & -56.252 & 3.987 & 207147688 & 11.76 & 10.83 & NO\\
J0309.9-6941 & J031115.2-693914 & 47.814 & -69.654 & 11.870 & 650295508 & 19.79 & 4.72 & NO\\
J0347.0-6400 & J034616.8-635906 & 56.570 & -63.985 & 2.390 & 237917969 & 17.06 & 15.33 & NO\\
J0347.5+0722 & J034741.1+072307 & 56.922 & 7.385 & 3.440 & 468890586 & 15.74 & 29.69 & NO\\
J0409.4-3822 & J040923.1-382035 & 62.346 & -38.343 & 2.678 & 160142146 & 11.17 & 21.58 & NO\\
J0413.4-8609 & J041049.8-861142 & 62.708 & -86.195 & 4.421 & 410365337 & 8.21 & 2.43 & NO\\
J0414.7-4300 & J041438.5-430123 & 63.661 & -43.023 & 4.145 & 685921034 & 19.68 & 27.22 & NO \\
J0506.6-4629 & J050636.4-462612 & 76.652 & -46.437 & 4.852 & 200326143 & 18.18 & 6.70 & NO\\
J0530.0-6900e & J052501.3-693832 & 81.255 & -69.642 & 7.062 & 729768454 & 15.18 & 21.38 & NO\\
J0553.9-5048 & J055359.5-505144 & 88.498 & -50.862 & 0.822 & 734657073 & 18.39 & 1.16 & NO\\
J0600.7+2012c & J060147.0+200923 & 90.446 & 20.157 & 5.372 & 429496397 & 11.37 & 14.23 & YES\\
J0608.6-2305 & J060854.6-230749 & 92.228 & -23.130 & 4.401 & 124182419 & 14.91 & 7.43 & NO\\
J0614.1+1628c & J061252.8+163631 & 93.220 & 16.609 & 3.940 & 294361956 & 13.15 & 7.46 & NO\\
J0616.5+2235 & J061637.6+223630 & 94.157 & 22.608 & 10.457 & 426519613 & 16.47 & 27.52 & NO\\
J0618.7+1211 & J061816.9+122031 & 94.571 & 12.342 & 3.357 & 715475807 & 10.03 & 0.62 & YES\\
J0622.5+3120 & J062254.9+312035 & 95.729 & 31.343 & 3.530 & 10000410370 & 16.16 & 5.74 & NO\\
J0624.7-4903 & J062423.4-490211 & 96.098 & -49.036 & 2.445 & 255562792 & 18.18 & 25.40 & NO\\
J0634.6-3046c & J063350.4-305309 & 98.460 & -30.886 & 4.114 & 51995003 & 14.76 & 12.15 & NO\\
J0639.1-8009 & J064100.6-801127 & 100.253 & -80.191 & 1.680 & 764458824 & 18.08 & 4.83 & uncertain\\
J0648.6-3623 & J064759.3-362631 & 101.997 & -36.442 & 5.229 & 148343363 & 17.79 & 19.41 & NO\\
J0650.2-5144 & J065009.9-514418 & 102.541 & -51.738 & 5.196 & 238003096 & 17.68 & 4.97 & NO\\
J0653.8-1330 & J065349.1-132933 & 103.455 & -13.493 & 6.711 & 142668337 & 14.43 & 27.89 & NO\\
J0700.8+1551 & J070045.1+154839 & 105.188 & 15.811 & 3.674 & 386487337 & 18.01 & 28.12 & NO\\
J0704.9-1007c & J070512.2-100749 & 106.301 & -10.130 & 4.489 & 177122224 & 12.03 & 5.60 & YES \\
J0706.8-4351 & J070556.6-435629 & 106.486 & -43.941 & 3.817 & 770871882 & 17.17 & 27.87 & NO\\
J0712.8-5637 & J071432.5-563947 & 108.636 & -56.663 & 3.499 & 294276020 & 18.28 & 17.37 & NO\\
J0715.7-1128 & J071454.6-113345 & 108.728 & -11.563 & 4.803 & 178642593 & 17.25 & 16.63 & NO\\
J0722.4-2650 & J072219.3-264734 & 110.581 & -26.793 & 3.510 & 107002136 & 11.93 & 4.13 & NO\\
J0725.3-2550c & J072620.1-254331 & 111.584 & -25.725 & 4.903 & 107858606 & 12.69 & 4.55 & YES\\
J0725.6-5008 & J072540.0-500656 & 111.417 & -50.116 & 3.331 & 262492759 & 14.88 & 7.32 & NO\\
J0725.7-0549 & J072547.6-054826 & 111.449 & -5.807 & 2.438 & 753769992 & 18.44 & 24.86 & NO\\
J0752.0-2931 & J075210.2-293018 & 118.043 & -29.505 & 3.081 & 776193304 & 12.40 & 3.48 & YES\\
J0753.1-2624c & J075219.7-262555 & 118.082 & -26.432 & 4.095 & 128565725 & 8.93 & 10.99 & uncertain\\
J0757.9-1514 & J075813.7-151532 & 119.557 & -15.259 & 3.488 & 784138896 & 11.28 & 27.64 & NO\\
J0800.9+0733 & J080056.3+073237 & 120.235 & 7.544 & 6.341 & 387975749 & 16.57 & 20.71 & NO\\
J0807.2+3312 & J080658.8+331321 & 121.745 & 33.223 & 5.226 & 10001614445 & 15.92 & 16.60 & NO\\
J0824.1-6004 & J082417.5-600430 & 126.073 & -60.075 & 3.505 & 309709714 & 17.17 & 27.74 & NO\\
J0825.6-5216c & J082632.0-523128 & 126.634 & -52.525 & 4.324 & 140149342 & 16.75 & 15.95 & NO\\
J0829.3+3729 & J082944.7+373027 & 127.436 & 37.508 & 4.398 & 307843917 & 12.58 & 3.57 & NO\\
J0830.1-5454c & J083030.9-545302 & 127.629 & -54.884 & 3.564 & 89709755 & 12.83 & 3.40 & YES\\
J0837.8-4048c & J083811.6-404445 & 129.549 & -40.746 & 4.684 & 180737955 & 16.60 & 4.93 & NO\\
J0838.4-3952 & J083743.9-395318 & 129.433 & -39.889 & 2.139 & 180730456 & 10.75 & 17.77 & YES\\
J0838.9-2502 & J083832.5-250226 & 129.636 & -25.041 & 3.587 & 196508303 & 8.71 & 3.64 & uncertain\\
J0843.9-4224c & J084309.7-422112 & 130.791 & -42.354 & 11.561 & 430426680 & 11.90 & 19.19 & NO\\
J0844.9-4117 & J084545.8-411033 & 131.441 & -41.176 & 2.870 & 181509708 & 11.73 & 2.02 & NO\\
J0847.8-4138 & J084742.1-414412 & 131.926 & -41.737 & 3.797 & 181741837 & 5.83 & 19.84 & NO\\
J0850.3-4448 & J085014.1-445506 & 132.559 & -44.918 & 6.866 & 29125381 & 14.71 & 21.57 & NO\\
J0853.2-4218c & J085253.6-421357 & 133.224 & -42.233 & 13.295 & 29796236 & 11.43 & 20.30 & NO\\
J0859.2-4729 & J085905.4-473042 & 134.773 & -47.512 & 2.082 & 10000767461 & 10.95 & 28.29 & NO\\
J0859.3-4342 & J085922.5-434516 & 134.844 & -43.755 & 2.745 & 10000767435 & 10.82 & 17.17 & NO\\
J0900.1-4402c & J090002.5-440616 & 135.011 & -44.104 & 4.363 & 30966771 & 13.55 & 22.49 & NO\\
J0906.8-2122 & J090655.4-211636 & 136.731 & -21.277 & 5.383 & 432083724 & 13.26 & 6.27 & NO\\
J0910.1-1816 & J091004.0-181615 & 137.517 & -18.271 & 3.158 & 432139531 & 15.77 & 13.37 & NO\\
J0917.9-4755 & J091859.9-475800 & 139.750 & -47.967 & 1.549 & 75984116 & 6.16 & 4.56 & NO\\
J0919.5-6203 & J092001.8-620122 & 140.008 & -62.023 & 2.656 & 359159167 & 17.02 & 24.85 & NO\\
J0942.3-3530 & J094128.6-352827 & 145.369 & -35.474 & 3.866 & 873136917 & 16.11 & 12.64 & NO\\
J0944.3-0911 & J094424.4-091051 & 146.102 & -9.181 & 3.538 & 96029489 & 16.10 & 14.35 & NO\\
J0944.7-5124 & J094434.1-512316 & 146.142 & -51.388 & 5.323 & 363313927 & 13.65 & 24.24 & NO\\
J0947.0-3548 & J094614.4-355226 & 146.560 & -35.874 & 4.847 & 12008346 & 13.27 & 5.86 & NO\\
J0951.8-5944 & J095112.9-594701 & 147.804 & -59.784 & 2.813 & 358275127 & 9.72 & 4.34 & NO\\
J0959.1-6441 & J095930.3-644003 & 149.876 & -64.668 & 1.791 & 374227719 & 10.85 & 1.85 & YES\\
J1000.5-5709c & J095955.0-570805 & 149.979 & -57.135 & 2.561 & 461974137 & 12.99 & 2.03 & NO\\
J1000.7-4248 & J100007.0-424232 & 150.029 & -42.709 & 5.097 & 36330893 & 15.96 & 25.83 & NO\\
J1011.4-5729c & J101102.4-572850 & 152.760 & -57.481 & 10.326 & 463036286 & 10.70 & 15.03 & uncertain\\
J1016.1-4247 & J101620.6-424722 & 154.086 & -42.790 & 2.607 & 102123684 & 15.84 & 27.13 & NO\\
J1020.4-5314 & J102001.5-531341 & 155.006 & -53.228 & 5.386 & 220252477 & 13.07 & 10.08 & NO\\
J1036.3-5833e & J102556.2-574843 & 156.485 & -57.812 & 0.900 & 464570167 & 10.56 & 1.70 & NO \\
J1036.5-7434c & J103417.1-741442 & 158.571 & -74.245 & 0.977 & 453740197 & 14.65 & 22.53 & NO\\
J1054.0-5938 & J105432.8-593056 & 163.637 & -59.516 & 5.103 & 459819162 & 10.08 & 23.57 & YES\\
J1057.6-4051 & J105729.9-405108 & 164.375 & -40.852 & 2.519 & 158548040 & 15.84 & 15.06 & NO\\
J1058.9-6101 & J105900.3-610850 & 164.751 & -61.147 & 2.946 & 465944979 & 11.10 & 3.50 & NO\\
J1109.4-6115e & J110146.7-610125 & 165.445 & -61.024 & 3.011 & 466289675 & 9.94 & 24.99 & NO\\
J1110.4-8023 & J111044.9-802135 & 167.687 & -80.360 & 1.179 & 395014156 & 12.78 & 29.74 & NO\\
J1127.9-6158 & J112753.9-620125 & 171.975 & -62.024 & 2.002 & 316656895 & 10.28 & 2.37 & NO\\
J1133.7-6223c & J113331.2-621931 & 173.380 & -62.325 & 6.616 & 318676471 & 13.61 & 4.70 & YES\\
J1138.4-3555 & J113851.1-355246 & 174.713 & -35.880 & 5.151 & 181206990 & 18.92 & 28.39 & NO\\
J1146.0-0638 & J114600.6-063854 & 176.503 & -6.649 & 4.589 & 144145759 & 17.46 & 26.75 & NO\\
J1154.5-5952 & J115422.7-595815 & 178.595 & -59.971 & 5.764 & 305625824 & 9.70 & 4.64 & NO\\
J1155.2-1111 & J115514.9-111121 & 178.812 & -11.189 & 4.407 & 902390769 & 19.55 & 2.15 & NO\\
J1203.5-5745 & J120237.3-573808 & 180.656 & -57.636 & 3.408 & 376553667 & 14.27 & 12.20 & NO\\
J1203.7-6303c & J120347.0-630352 & 180.946 & -63.065 & 3.776 & 379340421 & 12.01 & 29.59 & YES\\
J1204.3-6111 & J120507.4-610557 & 181.281 & -61.099 & 1.529 & 379547065 & 10.12 & 2.33 & NO\\
J1206.8-5836 & J120613.2-583330 & 181.555 & -58.559 & 3.538 & 67014959 & 13.06 & 24.41 & NO\\
J1206.8-6038c & J120616.6-604136 & 181.570 & -60.694 & 3.842 & 380121855 & 11.69 & 5.75 & YES\\
J1209.2-6009 & J120753.8-601034 & 181.974 & -60.176 & 5.130 & 380801636 & 13.50 & 27.28 & NO\\
J1210.4-6250 & J120958.5-624958 & 182.494 & -62.833 & 4.360 & 381373887 & 8.98 & 26.18 & NO\\
J1213.6-5954 & J121346.3-595025 & 183.443 & -59.840 & 3.922 & 411538281 & 15.13 & 16.30 & NO\\
J1217.2-2500 & J121734.6-245514 & 184.394 & -24.921 & 4.144 & 204641269 & 17.01 & 24.88 & NO\\
J1231.7-6653 & J123232.9-665917 & 188.137 & -66.988 & 3.171 & 326291208 & 13.00 & 25.53 & NO\\
J1303.1-4714 & J130311.8-471117 & 195.799 & -47.188 & 2.800 & 359690876 & 13.32 & 1.79 & NO\\
J1309.1-6223 & J130855.3-622442 & 197.231 & -62.412 & 2.168 & 441485051 & 9.92 & 20.25 & NO\\
J1312.6-6231c & J131229.1-623431 & 198.122 & -62.575 & 5.187 & 442382268 & 10.92 & 4.42 & uncertain\\
J1320.3-6410c & J132015.5-641352 & 200.065 & -64.231 & 2.238 & 449410766 & 12.38 & 6.97 & NO\\
J1321.1-6239 & J132139.2-623304 & 200.414 & -62.551 & 4.662 & 449549266 & 13.45 & 7.92 & uncertain\\
J1325.3-5413 & J132528.1-541137 & 201.367 & -54.194 & 5.154 & 359394835 & 14.12 & 11.17 & NO\\
J1325.4-4706 & J132539.7-470329 & 201.415 & -47.058 & 4.239 & 1048224485 & 17.26 & 22.07 & NO\\
J1329.9-6108 & J133010.7-610713 & 202.545 & -61.121 & 6.529 & 314725885 & 10.57 & 6.53 & NO\\
J1335.5-4546 & J133512.9-454833 & 203.804 & -45.809 & 4.790 & 243261572 & 16.12 & 9.23 & NO\\
J1336.9-4611 & J133648.7-461648 & 204.203 & -46.280 & 3.768 & 243294498 & 15.07 & 27.60 & NO\\
J1350.9-2757 & J134954.3-281110 & 207.477 & -28.186 & 1.780 & 441835320 & 7.29 & 1.69 & NO\\
J1356.0-6747 & J135739.0-673707 & 209.413 & -67.619 & 5.236 & 448295110 & 10.29 & 2.06 & NO\\
J1357.3-6123 & J135657.6-612317 & 209.240 & -61.388 & 2.060 & 324566752 & 11.82 & 4.43 & NO\\
J1357.8-5724 & J135728.7-572759 & 209.370 & -57.467 & 4.276 & 208873139 & 12.59 & 2.94 & NO\\
J1358.0-7749 & J135832.7-774528 & 209.636 & -77.758 & 4.768 & 1003671126 & 17.74 & 28.67 & NO\\
J1407.7-3017 & J140806.8-302354 & 212.029 & -30.399 & 0.526 & 272152804 & 15.43 & 1.55 & NO\\
J1409.1-6121e & J140513.5-610734 & 211.307 & -61.126 & 4.287 & 329221764 & 12.94 & 3.46 & NO\\
J1418.2-5400 & J141658.3-535551 & 214.243 & -53.931 & 1.118 & 331335557 & 9.02 & 3.29 & NO\\
J1424.8-8012 & J142344.7-800720 & 215.936 & -80.122 & 1.757 & 282331633 & 12.01 & 2.88 & NO\\
J1429.8-0739 & J142949.5-073307 & 217.457 & -7.552 & 2.239 & 10000760650 & 16.04 & 3.41 & NO\\
J1431.5-6627 & J143211.2-663531 & 218.047 & -66.592 & 4.163 & 446745475 & 13.96 & 14.31 & NO\\
J1441.4-1934 & J144128.0-193553 & 220.367 & -19.598 & 1.331 & 1179126750 & 17.25 & 26.61 & NO\\
J1442.1-2800 & J144157.0-280359 & 220.488 & -28.067 & 3.677 & 52357 & 5.75 & 3.34 & YES\\
J1451.6-3726 & J145132.2-372943 & 222.884 & -37.495 & 3.990 & 160341056 & 14.93 & 9.01 & NO\\
J1454.3-5551 & J145546.4-555835 & 223.944 & -55.977 & 5.912 & 419165490 & 10.92 & 7.23 & NO\\
J1456.3-5419 & J145615.1-542205 & 224.063 & -54.368 & 5.462 & 419482394 & 14.56 & 12.28 & NO\\
J1505.1-5145 & J150402.5-514601 & 226.011 & -51.767 & 5.537 & 370300045 & 15.02 & 2.94 & NO\\
J1508.4-4817 & J150805.1-480959 & 227.021 & -48.166 & 2.579 & 120463113 & 6.39 & 1.32 & NO\\
J1511.2-5803 & J151137.7-580113 & 227.907 & -58.020 & 3.787 & 41160704 & 10.74 & 4.81 & NO\\
J1513.0-3118 & J151244.5-311651 & 228.186 & -31.281 & 1.290 & 371122327 & 10.64 & 3.07 & NO\\
J1516.4-2640 & J151605.2-264726 & 229.022 & -26.791 & 2.921 & 61596933 & 12.67 & 4.05 & uncertain\\
J1517.0-4600 & J151735.9-460756 & 229.400 & -46.132 & 5.348 & 144026645 & 13.39 & 10.42 & NO\\
J1517.7-4446 & J151728.3-444253 & 229.368 & -44.715 & 4.338 & 143421375 & 12.98 & 27.45 & NO\\
J1527.9-4943 & J152824.6-494120 & 232.103 & -49.689 & 5.156 & 143614965 & 13.36 & 1.33 & NO\\
J1529.4-6027 & J152953.8-603616 & 232.474 & -60.605 & 2.821 & 461142177 & 12.24 & 4.07 & YES\\
J1533.0-6239 & J153312.7-623248 & 233.303 & -62.547 & 2.493 & 271738439 & 12.94 & 10.96 & YES\\
J1534.4-6719 & J153451.4-672105 & 233.714 & -67.352 & 2.826 & 446999814 & 12.93 & 2.01 & NO\\
J1545.2-4553 & J154519.1-455412 & 236.330 & -45.903 & 4.007 & 254621615 & 12.48 & 5.56 & NO\\
J1547.4-4802 & J154725.1-480755 & 236.855 & -48.132 & 2.227 & 270585530 & 11.81 & 4.45 & YES\\
J1551.9-6015 & J155033.8-601409 & 237.641 & -60.236 & 6.527 & 339854307 & 13.38 & 22.34 & NO\\
J1553.8-5325e & J155238.5-532612 & 238.161 & -53.437 & 1.267 & 281008212 & 6.83 & 2.18 & NO\\
J1604.4-4927 & J160412.6-492211 & 241.053 & -49.370 & 4.355 & 214760610 & 11.09 & 5.86 & NO\\
J1609.3-4326 & J160942.8-433005 & 242.429 & -43.502 & 3.080 & 163011203 & 13.53 & 6.72 & NO\\
J1611.9-5125c & J161200.3-512529 & 243.002 & -51.425 & 3.741 & 216057258 & 9.97 & 6.77 & NO\\
J1616.6-5341 & J161648.6-534141 & 244.203 & -53.695 & 4.045 & 411191632 & 12.15 & 1.28 & NO\\
J1620.8-5035c & J162024.1-503710 & 245.101 & -50.620 & 2.712 & 316549457 & 12.96 & 2.90 & uncertain\\
J1622.7-4934c & J162231.7-493623 & 245.632 & -49.607 & 3.674 & 216851596 & 14.06 & 16.36 & NO\\
J1623.7-2315c & J162334.0-231747 & 245.892 & -23.296 & 2.295 & 203239672 & 16.98 & 19.00 & NO\\
J1624.9-4538 & J162540.6-454005 & 246.420 & -45.668 & 3.150 & 223995009 & 12.44 & 4.76 & NO\\
J1626.6-4251 & J162654.5-425222 & 246.727 & -42.873 & 3.716 & 224784564 & 13.90 & 1.31 & NO\\
J1633.4-5845 & J163339.4-582855 & 248.414 & -58.482 & 4.027 & 221580187 & 11.87 & 4.24 & NO\\
J1635.4-3249 & J163544.1-324939 & 248.934 & -32.828 & 2.858 & 280604073 & 14.96 & 26.30 & NO\\
J1635.4-3453 & J163520.4-344853 & 248.835 & -34.815 & 0.859 & 280361015 & 9.80 & 0.96 & NO\\
J1641.4-4803c & J164127.6-475605 & 250.365 & -47.935 & 4.913 & 232769736 & 11.51 & 3.56 & NO\\
J1649.3-4441 & J164921.7-444358 & 252.341 & -44.733 & 3.290 & 245914739 & 11.73 & 27.48 & YES\\
J1651.1-5848 & J165105.7-585854 & 252.774 & -58.982 & 3.544 & 170043698 & 12.61 & 24.45 & NO\\
J1652.2-4516 & J165154.3-451449 & 252.977 & -45.247 & 3.502 & 246541078 & 12.60 & 1.14 & NO\\
J1652.2-4633e & J164825.9-462209 & 252.108 & -46.369 & 3.487 & 245779952 & 12.66 & 9.19 & NO\\
J1654.5-5509 & J165610.6-553206 & 254.044 & -55.535 & 2.369 & 170861437 & 12.05 & 25.72 & NO\\
J1655.9-3101 & J165619.8-305935 & 254.083 & -30.993 & 4.542 & 35398332 & 13.58 & 27.61 & NO\\
J1705.1-4748 & J170500.2-475224 & 256.251 & -47.874 & 3.414 & 124473650 & 11.18 & 2.45 & YES\\
J1706.2-4950 & J170639.7-494914 & 256.666 & -49.821 & 4.757 & 212887369 & 12.70 & 6.44 & NO\\
J1706.4-4649c & J170734.9-464048 & 256.896 & -46.680 & 1.588 & 125958765 & 6.61 & 2.71 & YES\\
J1706.5-4023c & J170634.0-402543 & 256.642 & -40.429 & 3.004 & 377852306 & 12.46 & 5.11 & YES\\
J1717.6-4404 & J171729.0-440224 & 259.371 & -44.040 & 2.702 & 216810034 & 15.00 & 24.04 & NO\\
J1722.1-3205 & J172225.3-320503 & 260.606 & -32.084 & 4.926 & 156668851 & 10.34 & 28.70 & NO\\
J1722.1-4007 & J172226.3-400537 & 260.610 & -40.094 & 3.247 & 198063204 & 8.75 & 6.99 & NO\\
J1723.1-2859 & J172302.4-285157 & 260.760 & -28.866 & 2.501 & 85062000 & 11.74 & 5.49 & NO\\
J1730.1-4343 & J172959.6-433308 & 262.498 & -43.552 & 1.670 & 218367096 & 8.00 & 3.02 & NO\\
J1737.3-3332 & J173733.5-333236 & 264.390 & -33.544 & 4.775 & 105192617 & 13.03 & 15.48 & YES\\
J1739.5-2929 & J173920.5-293151 & 264.836 & -29.531 & 2.856 & 201343715 & 5.90 & 1.59 & NO\\
J1740.6-3430 & J173953.0-344118 & 264.971 & -34.688 & 3.885 & 107057795 & 11.30 & 2.41 & NO\\
J1740.7-6750 & J174044.8-674325 & 265.187 & -67.724 & 3.511 & 294292450 & 16.48 & 18.80 & NO\\
J1743.4-3406 & J174349.1-335657 & 265.955 & -33.949 & 3.384 & 109498376 & 9.65 & 1.43 & NO\\
J1744.9-3322 & J174516.3-331845 & 266.318 & -33.313 & 4.695 & 110879048 & 9.00 & 2.85 & NO\\
J1758.5-3219 & J175834.5-324059 & 269.644 & -32.683 & 3.339 & 261914765 & 11.10 & 24.77 & NO\\
J1829.6-4310 & J183000.8-430535 & 277.503 & -43.093 & 3.020 & 90999580 & 11.56 & 2.58 & NO\\
J1854.3-3640 & J185410.1-363923 & 283.542 & -36.657 & 11.641 & 253864764 & 8.87 & 18.30 & NO\\
J1904.0-5925 & J190438.7-592504 & 286.162 & -59.418 & 5.103 & 344030745 & 12.51 & 25.22 & NO\\
J1905.2-5120 & J190509.8-511943 & 286.291 & -51.329 & 7.508 & 7125649 & 11.85 & 10.86 & uncertain\\
J1910.1-4704 & J191058.3-470930 & 287.743 & -47.158 & 3.474 & 61383325 & 14.04 & 28.03 & NO\\
J1918.6-7813c & J192735.4-781309 & 291.898 & -78.219 & 3.812 & 257573215 & 8.07 & 6.24 & NO\\
J2013.9-8717 & J201511.7-871708 & 303.799 & -87.286 & 5.074 & 318299718 & 13.96 & 4.47 & uncertain\\
J2055.0-5218 & J205445.0-520948 & 313.688 & -52.163 & 4.700 & 1989670921 & 19.08 & 26.32 & NO\\
J2217.0-6727 & J221659.2-672757 & 334.247 & -67.466 & 1.877 & 261421261 & 18.70 & 3.08 & NO\\
J2241.4-8327 & J224157.5-832749 & 340.490 & -83.464 & 4.066 & 273733637 & 8.56 & 23.26 & NO\\
J2343.0-4756 & J234310.4-475356 & 355.794 & -47.899 & 5.621 & 2055562336 & 19.09 & 15.66 & NO\\
J2355.5-6614 & J235537.2-661253 & 358.905 & -66.215 & 4.411 & 267062030 & 17.20 & 22.32 & NO

\end{longtable}

\end{footnotesize}

\label{lastpage}

\end{document}